\documentclass[prd,aps,twocolumn,floatfix,superscriptaddress,preprintnumbers,nofootinbib]{revtex4-1}
\usepackage{amsfonts}
\usepackage{amssymb}
\usepackage{amsmath,mathtools}
\usepackage{color}

\RequirePackage{ifpdf}
\ifpdf
  \usepackage[pdftex]{graphicx}
\else
  \usepackage[dvipdfmx]{graphicx}
\fi

\newcommand{\braket}[1]{{\langle #1 \rangle}}

\newcommand{\tr}{\mathrm{tr}}

\begin{document}
\preprint{OU-HET-1093}

\title{Nuclear binding energy in holographic QCD}
\author{Koji Hashimoto}
\email{koji@scphys.kyoto-u.ac.jp}
\author{Yoshinori Matsuo}
\email{ymatsuo@gauge.scphys.kyoto-u.ac.jp}
\affiliation{Department of Physics, Osaka University, Toyonaka, Osaka 560-0043, Japan}
\affiliation{Department of Physics, Kyoto University, Kyoto 606-8502, Japan}
\date{\today}

\begin{abstract}
Saturation of the nuclear binding energy is one of the most important properties of atomic nuclei. We derive the saturation in holographic QCD, by building a shell-model-like mean-field nuclear potential from the nuclear density profile obtained in a holographic multi-baryon effective theory. 
The numerically estimated binding energy is close to the experimental value. 
\end{abstract}


\maketitle

\setcounter{footnote}{0}

\noindent

\tableofcontents

\section{Introduction and summary}
\label{sec1}

Holographic QCD 
provides effective theories via the AdS/CFT correspondence 
\cite{Maldacena:1997re,Gubser:1998bc,Witten:1998qj}
and allows us to calculate various observables
of large $N_c$ QCD-like gauge theories. 
Nuclear physics is a challenging target of holographic QCD.
The nuclear matrix model \cite{Hashimoto:2010je} 
was proposed to be a unified model of nuclear physics 
inspired and derived by the AdS/CFT. 
Atomic nuclei are bound states of nucleons,  
which may allow an effective description in terms of baryons in holographic QCD.
Nucleons are D-branes in the gravity side of the AdS/CFT, so the holographic theory
of the nucleons is given by matrices, which is the basis of
the nuclear matrix model. 
By the nuclear matrix model,
several important properties of nuclei, including the 
nuclear radii \cite{Hashimoto:2011nm}, nuclear spectra and magic numbers \cite{Hashimoto:2019wmg} 
have been reproduced.\footnote{
As for the nuclear matrix model, see \cite{Aoki:2012th} for a three-flavor case, \cite{Hashimoto:2010ue,Hashimoto:2009as} for the three-body force, and \cite{Hashimoto:2010rb} for the fermionic nature of nucleon.
An earlier proposal of the matrix description includes \cite{Hashimoto:2008jq,Hashimoto:2009pe}. 
Note that the binding of nucleons by nuclear force \cite{Hashimoto:2008zw,Hashimoto:2009ys,Kim:2009sr,Kim:2008iy,Cherman:2011ve}
may be problematic generically in holographic QCD 
\cite{Kaplunovsky:2010eh}.
For related approaches, see  \cite{Pomarol:2008aa,Pahlavani:2010zzb,Pahlavani:2014dma,Baldino:2017mqq,Bolognesi:2013jba,Kim:2007zm,Rozali:2007rx,Kim:2007vd,Rho:2009ym,Kaplunovsky:2012gb,Kaplunovsky:2015zsa,Jarvinen:2020xjh,Li:2015kma,Li:2016kfo,Baldino:2021uie}.
} 
In this paper, we study another important aspect of nuclei---the nuclear binding energy. 
We show that 
the nuclear matrix model also reproduces 
the saturation of the nuclear binding energy:
the property that 
the nuclear binding energy per nucleon approaches a constant 
independent of the mass number $A$
at large $A$.

Nuclear states in holographic QCD are obtained as energy eigenstates of 
the nuclear matrix model. 
In the gravity side of the Sakai-Sugimoto model \cite{Sakai:2004cn,Sakai:2005yt}, the baryons correspond to 
the D4-branes called baryon vertices which are wrapped on the $S^4$-directions 
in the Witten geometry.\footnote{
Baryon vertices are most stable when they are located at the tip of the Witten geometry
where the flavor D8-branes are placed, so the baryon vertices are on the flavor D8-branes. 
For simplicity, we ignore fluctuations in the directions away from the tip,
and the couplings to the gravity modes (as they are sub-leading in the large $N_c$ expansion).
}
The nuclear matrix model action is the low energy effective action of the baryon vertices. 
After the dimensional reduction for $S^4$-directions, 
the Hamiltonian of the nuclear matrix model for $A$ baryons is derived as \cite{Hashimoto:2010je}
\begin{align}
 H_\text{full} &= H - \tr A_t \left(Q_{U(A)} - N_c \right) \ , 
\\
 H
&=
 \frac{1}{2} \tr (\Pi^I)^2 + 2 \bar\pi_{\dot \alpha i}^a \pi^{\dot \alpha i}_a 
 + \frac{1}{2} M^2 \bar w^{\dot \alpha i}_a w_{\dot \alpha i}^a 
\notag\\
&\quad
  - 4 i \lambda \epsilon^{IJK} X_A^J X_B^K f^{AB}{}_C\, 
    \bar w^{\dot \alpha i}_a (\tau^I)_{\dot \alpha}{}^{\dot\beta} (t^C)^a{}_b w_{\dot \beta i}^b 
\notag\\
&\quad
  - 2 \lambda \tr \left[X^I , X^J \right]^2
  + \lambda \left(\bar w^{\dot \alpha i} (\tau^I)_{\dot \alpha}{}^{\dot\beta} w_{\dot \beta i}\right)^2 \ , 
\label{H}
\end{align}
where the indices $\{I,J,K,\cdots\}$, $\{i,j,k,\cdots\}$ and 
$\{\dot \alpha,\dot \beta,\dot \gamma,\cdots\}$ label 
the three dimensional space, $SU(N_f)$ flavors and $SU(2)$ spins, respectively. 
The other indices $\{A,B,C,\cdots\}$ and $\{a,b,c,\cdots\}$ stand for 
the adjoint and fundamental representations of the baryon $U(A)$ symmetry, 
and the trace is taken over the baryon $U(A)$ matrices. 
The fields $X^I$, $w$ and $\bar w$ are scalar fields 
which come from the D4-D4 and D4-D8 open strings, and 
$\Pi^I$, $\pi$ and $\bar \pi$ are their conjugate momenta, respectively. 

The nuclear matrix model \eqref{H} gives 
bound states of nucleons and reproduces nuclear states \cite{Hashimoto:2019wmg}.
There is a  difficulty in directly 
evaluating the nuclear binding energy by the nuclear matrix model. 
The nuclear binding energy is the energy difference between the nuclear bound state and 
the state at which all the constituent nucleons are infinitely far apart, while the nuclear
matrix model is effective for nucleons close to each other.%
\footnote{
When the eigenvalues 
of $X$ are far away, flat directions should appear for which supersymmetry restoration would be important. 
The nuclear matrix model ignores fermions on the baryon vertex.
}
This difficulty is overcome once the following point is noticed: 
the saturation of the binding energy is only for large $A$.
Most of the total binding energy comes from nucleons in lower energy levels, 
and that of the highest energy level which should be compared to the nucleon separated far away
can be treated as zero approximately. 
Therefore, we can calculate the binding energy of the other nucleons and then the total binding energy 
by solving the nuclear matrix model. 

Nuclear states in the matrix model are calculated in \cite{Hashimoto:2019wmg} for small baryon numbers, 
but the procedure is not useful for large nucleon numbers. 
In the large $A$ limit, it is convenient to use the mean-field approximation. 
We here adopt the following strategy. 
First, starting with the nuclear density profile obtained by the nuclear matrix model \cite{Hashimoto:2011nm}, 
we 
derive an effective mean-field potential which reproduces the density profile. 
This potential defines a {\it holographic nuclear shell model}, 
in which 
positions of nucleons are identified with
diagonal eigenvalues of $X$
of the nuclear matrix model.%
\footnote{Eigenvalue were shown to behave as fermions
when $N_c$ is odd \cite{Hashimoto:2010rb}.} 
Assuming that the binding energy for a nucleon 
at 
the fermi level is negligible, 
the total binding energy of the nucleus can be obtained by 
summing up all the energy of nucleons below the fermi energy 
in the holographic nuclear shell model. 
We calculate this nuclear binding energy and show that it reproduces the saturation property. 

Since the nuclear matrix model can treat 
flavors, spins and orbital motion of baryons in a unified fashion,
we can estimate the magnitude of the nuclear binding energy using
the nucleon mass and the $\Delta$ mass as inputs. We find that the resultant numerical value turns out
to be close to 
the experimental value of the binding energy. 
The result is quite nontrivial 
in view of the crude approximations employed in the holographic QCD. 

The organization of this paper is as follows. 
In Sec.~\ref{sec2}, 
we construct a holographic nuclear shell model by deriving 
an effective potential from the nuclear density profile 
in the nuclear matrix model.
In Sec.~\ref{sec4}, we calculate the holographic nuclear binding energy. 
Using the holographic nuclear shell model, 
we obtain 
the nuclear binding energy per nucleon 
and show that it is independent of $A$. 
We find that the binding energy is numerically close to 
the experimental value. Appendix A describes the details of the mass rescaling used in the calculus, 
and Appendix B adds a novel observation on the $A$-dependence of nuclear radii.

\section{Holographic nuclear shell model}
\label{sec2}

First, we review the nuclear density profile of the 
nuclear matrix model \cite{Hashimoto:2011nm} and check its consistency by confirming that
the effect of $w$ is sub-leading. 
Then in Subsec.~\ref{sec3} 
we inversely obtain the shell-model potential from the density profile.
This serves as a holographic nuclear shell model, with which in Sec.~\ref{sec4} we 
calculate the nuclear binding energy.

\subsection{Nuclear density in the nuclear matrix model}

The nuclear density profile in the nuclear matrix model was derived in \cite{Hashimoto:2011nm} 
by using 
the Ramond-Ramond charge density formula of D-branes \cite{Taylor:1999gq}, 
\begin{align}
 \rho(x) &= \frac{1}{(2\pi)^3} \int d^3 k \, e^{-ik_I x^I}
  \left\langle \tr \exp\left[ik_J X^J\right]\right\rangle \ . 
  \label{EV}
\end{align}
In the large $A$ and large $D$ limit, the nuclear matrix model effectively behaves as 
that with a harmonic potential since only the ladder diagrams contribute to the expectation value above. 
The density profile was calculated in \cite{Hashimoto:2011nm} as%
\footnote{
The evaluation of \eqref{EV} in the large $A$ limit and in the large $D$ limit 
(where $D$ refers to the index $I$ of $X^I$ as $I=1,2,\cdots, D$ and will be set to $D=3$ after the evaluation) is described in \cite{Hashimoto:2011nm}. 
At the leading order, the non-perturbative vacuum is found to give a non-zero expectation value for
$\langle {\rm tr}_A [X^I X^I]\rangle$, around which the fluctuation $X^I$ behaves as a massive free scalar. Then the evaluation of \eqref{EV} results in $\langle {\rm tr}_A \exp(ik\cdot X)\rangle 
\propto J_1(r_0 |k|)/(r_0 |k|)$ where $J_1$ is a Bessel function, whose inverse Fourier transform at $D=3$ provides \eqref{rho-mm}.
}
\begin{align}
 \rho(x) &= 
 \begin{cases}
  \displaystyle \frac{A}{\pi^2 r_0^2\sqrt{r_0^2 - r^2}} & (r<r_0)
  \\
  \rule{0pt}{18pt} 0 & (r>r_0)
 \end{cases}
 \label{rho-mm}
\end{align}
where $r$ is the radial coordinate in three dimensional space and 
$r_0$ is the surface radius 
which is related to 
the effective frequency $m$ as $r_0^2 = 2A/m$.
In the derivation, the $w$-sector was ignored 
because the number of degrees of freedom of $w$ is ${\cal O}(A)$ while that of $X$ is ${\cal O}(A^2)$. 
In this section we employ the ground state wave function developed in \cite{Hashimoto:2019wmg},
and find that indeed the effect of the $w$-sector is sub-leading, to make sure that 
we can use \eqref{rho-mm} in the subsequent sections.

Let us evaluate the total energy including the contribution from the $w$-sector.
The potential for $X^I$ can be approximated by 
a harmonic potential $(m^2/2)\tr (X^I)^2$ as in \cite{Hashimoto:2019wmg}
and the energy of the orbital motion becomes 
\begin{equation}
 E = m \left(N_X + \frac{3}{2}\left(A^2-1\right)\right) \ , 
\end{equation}
where $N_X$ is the number of excitations of $X^I$ and the second term comes from the zero-point fluctuations. 
The effective frequency of the orbital motion, $m$, is determined in a self-consistent fashion 
(see  Appendix B of \cite{Hashimoto:2019wmg}) and is approximately given by 
\begin{equation}
 m^2 = \frac{16 A \lambda}{3(A^2 -1)} \langle X^I_A X^I_A \rangle \ . 
\label{m<>}
\end{equation}
 
Here the expectation value in \eqref{m<>} is with the wave function given in \cite{Hashimoto:2019wmg},
and we have some remarks.
The gauge field $A_t$ behaves as a Lagrange multiplier 
and gives constraints that physical states must be in singlet of $SU(A)$ 
and have $U(1)$ charge of $Q_{U(1)}=N_c A$. 
Thus, the physical ground states must have $N_c A$ excitations of $w$. 
The energy of these excitations gives a correction to the nucleon mass, 
and also modifies the kinetic term of $X^I$ (see Appendix A for the details). 
Using an appropriate redefinition of $X^I$, the coupling constant $\lambda$ for $X^I$ 
in the Hamiltonian is replaced by 
\begin{equation}
 \lambda_r = \lambda \frac{M_0^3}{M_N^3} \ , 
 \label{lambdar}
\end{equation}
where $M_0$ is the bare tension of the baryon vertex and 
$M_N$ is the nucleon mass which includes the energy of $w$ and $\bar w$. 

As described in \cite{Hashimoto:2019wmg}, 
for $A > 2 N_f$, because it is impossible to form a totally antisymmetric combination solely of $w$,
excitations of $X^I$ should also be introduced in the wave function.\footnote{ 
Because of this fact, it was argued in \cite{Hashimoto:2019wmg} that 
the energy eigenstates of the nuclear matrix model
have a structure similar to those in the nuclear shell model. 
}
A straightforward counting shows that 
$N_X$ is approximately given by 
$ \left(3/2\right)^{7/3}A^{4/3}$
for large $A$. 
Thus, 
using the virial theorem $(1/2)m^2\langle \tr X^2\rangle = \braket{V} = E/2$, 
an estimation of $\langle \tr X^2 \rangle$ leads to
\begin{equation}
 \langle \tr X^2 \rangle 
 = \frac{1}{m}\left[\left(\frac{3}{2}\right)^{\!7/3}\!\!\!\!A^{4/3} + \frac{3}{2}(A^2-1)\right] \ . 
 \label{<X^2>}
\end{equation}
We find that, for large $A$, 
the first term which is the effect of the excitations 
due to the $w$-sector wave function 
is negligible (see Appendix B for the comparison  
of the sub-leading terms found here with the experimental data
of the nuclear radius).\footnote{
Note that in this paper we employ $N_c=3$.  If we took the large $N_c$ limit first, the first term in \eqref{<X^2>} would be dominant, as it is ${\cal O}(N_c^1)$
while the second term in ${\cal O}(N_c^0)$. 
} 
Thus, the effective frequency for large $A$ is evaluated as 
\begin{equation}
 m^3 = 8 \lambda_r A \ . 
\label{m^3}
\end{equation}
The density profile is given by \eqref{rho-mm}
with $r_0 = \lambda_r^{-1/6}A^{1/3}$.

\subsection{Derivation of mean-field potential}
\label{sec3}

In this subsection, 
we derive a 
mean-field effective potential for nucleons  
from the density distribution \eqref{rho-mm}. 
The wave function of a non-relativistic fermion 
in an arbitrary spherically symmetric potential $V(r)$ 
is given by using the WKB approximation as 
\begin{equation}
 \psi = \frac{C}{r} p^{-1/2}(r) e^{i\int dr\,p(r)} Y_{lm} \ , 
\end{equation}
where 
\begin{equation}
 p(r) = \sqrt{E-V(r)-\frac{l(l+1)}{r^2}} \ , 
\end{equation}
and $Y_{lm}$ is the spherical harmonics. 
The wave function is suppressed very fast outside the classical turning point 
and approximately zero for 
\begin{equation}
 E < V(r) + \frac{l(l+1)}{r^2} \ . 
\end{equation}
The wave function must satisfy the quantization condition, 
\begin{equation}
 \int dr\,p(r) \simeq \pi n \ , 
\end{equation}
with a positive integer $n$. 

The constant $C$ is fixed by the normalization condition as 
\begin{equation}
 C^{-2} = 4 \pi \int \frac{dr}{p(r)} \ . 
\end{equation}
When the fermions occupy all the states below the fermi level at $E = E_f$, 
the density is obtained by the sum of the probabilities for those states as 
\begin{align}
 \rho(r) 
 &= 4 \sum_{n,l,m} \left|\psi(r)\right|^2
 \notag\\
 &\simeq 
 \int_{V(r)}^{E_f} dE \int_0^{r^2(E-V(r))} (2l+1) dl\, \frac{dn}{dE} \frac{4 C^2}{r^2p(r)} \ , 
\end{align}
where the sum is approximated by an integral when the number of states is sufficiently large. 
The factor $4$ in the first line comes from the sum over spins and flavors (proton and neutron).
From the quantization condition, we obtain 
\begin{align}
 \frac{dn}{dE} 
 = 
 \frac{1}{\pi} \frac{d}{dE} \int dr \sqrt{E-V(r)-\frac{l(l+1)}{r^2}}
 = \frac{1}{8\pi^2 C^2} \ , 
\end{align}
and then, the density is calculated as 
\begin{align}
 \rho(r) 
 &= \frac{2}{3\pi^2}\left(E_f-V(r)\right)^{3/2} \ . 
\end{align}
Thus, for the nucleon density \eqref{rho-mm}, 
the effective potential is inversely given by 
\begin{equation}
 V(r) = E_f - \frac{3^{2/3}A^{2/3}}{2^{2/3} r_0^{4/3}\left(r_0^2-r^2\right)^{1/3}} \ , 
 \label{V}
\end{equation}
for $r<r_0$. 
Since the density is approximately zero for $r > r_0$, 
this effective potential must be accompanied by a steep potential barrier at $r = r_0$. 

The mean-field potential \eqref{V} defines our {\it holographic nuclear shell model}. In the next section,
we use this potential to evaluate the nuclear binding energy.

\section{Holographic nuclear binding energy}
\label{sec4}

\subsection{Saturation}

The shell-model effective potential \eqref{V} 
has the depth measured from the fermi level typically of 
\begin{equation}
 - V(r) \sim \frac{A^{2/3}}{r_0^2} \propto A^0 \ . 
\end{equation}
This roughly implies that the nuclear binding energy per nucleon is independent of $A$. 
In this section, we evaluate the nuclear binding energy by the effective potential \eqref{V}, 
more precisely. 

It is expected that the binding energy of nucleons at the fermi level is very small. 
Hence we take $E_f = 0$, and then, the binding energy per nucleon is approximately given by 
the average of the energy of each state. 
The total binding energy $B$ of a nucleus is a sum of the potential energy $B^{({\rm p})}$
and the kinetic energy $B^{({\rm k})}$, $B= B^{({\rm p})}+B^{({\rm k})}$. 
Below we calculate them separately. 

The total potential energy $B^{({\rm p})}$ is given in terms of the density as 
\begin{align}
 B^{({\rm p})} \equiv \left\langle V \right\rangle 
 &= 4\pi \int r^2 \rho(x) V(r) dr \ . 
\end{align}
Then, the total potential energy for the effective potential \eqref{V} is evaluated as 
\begin{align}
B^{({\rm p})}
 &= 
 - \int \frac{2^{4/3}\cdot 3^{2/3}A^{5/3} r^2 dr}{\pi r_0^{10/3}(r_0^2 - r^2)^{5/6}}
 \notag\\
 &= 
 - \frac{3^{8/3} \Gamma(\frac{7}{6})\,\lambda_r^{1/3} A}{2^{2/3}\sqrt{\pi}\,\Gamma(\frac{2}{3})} \ . 
\end{align}

By using the virial theorem, the total kinetic energy $B^{({\rm k})}$ is given by 
\begin{align}
B^{({\rm k})} \equiv  \frac{1}{2} \left\langle r\,V'(r) \right\rangle 
 &= 
 2\pi \int r^3 \rho(x) V'(r) dr \ . 
\end{align}
Since the potential has very high potential barrier at $r=r_0$, $V'(r)$ diverges there. 
In order to avoid this divergence, we introduce a regularization by modifying the potential as 
\begin{equation}
 V(r) = - \frac{3^{2/3}A^{2/3}}{2 r_0^{5/3}} \frac{(r_0 - r)^n}{\epsilon^{n + \frac{1}{3}}}
\end{equation}
for $r > r_0 - \epsilon$, and take the $\epsilon\to 0$ limit. 
Here, $n$ is the artifact of the regularization. 
It should be noted that the density should be modified simultaneously as 
\begin{equation}
 \rho(x) = \frac{A}{2^{1/2}\pi^2 r_0^{5/2}} 
 \frac{(r_0-r)^{ \frac{3}{2}n}}{\epsilon^{ \frac{3}{2}n + \frac{1}{2}}} \ , 
\end{equation}
for consistency with \eqref{rho-mm}. 
Then, the total kinetic energy is calculated as 
\begin{align}
B^{({\rm k})}
&  = 
 - \int_0^{r_0-\epsilon} \frac{2^{4/3} A^{5/3} r^4 dr}{3^{1/3}\pi^2 r_0^{10/3}(r_0^2 - r^2)^{11/6}}
\notag\\
&\qquad 
 - \int_{r_0 - \epsilon}^{r_0} 
 \frac{3^{2/3} A^{5/3} n (r_0 - r)^{ \frac{5}{2}n-1} dr}
 {2^{1/2}\pi r_0^{7/6} \epsilon^{ \frac{5}{2}n + \frac{5}{6}}}
\notag\\
 & = 
 \frac{3^{8/3}\sqrt{\pi}\,\lambda_r^{1/3} A}{2^{2/3}5\Gamma(\frac{2}{3})\Gamma(\frac{5}{6})} \ . 
\end{align}
The final expression is, in fact, independent of the artifact of the regularization. 

Thus, all together, the holographic nuclear binding energy $B$ in the large $A$ limit is obtained as 
\begin{equation}
\frac{B}{A}=
 \left[\frac{3^{8/3} \Gamma(\frac{7}{6})}{2^{2/3}\sqrt{\pi}\,\Gamma(\frac{2}{3})}
 -\frac{3^{8/3}\sqrt{\pi}}{2^{2/3}5\Gamma(\frac{2}{3})\Gamma(\frac{5}{6})}\right] \lambda_r^{1/3} \ .
 \label{binres}
\end{equation}
The important point is that this is independent of $A$, 
which shows the saturation of the nuclear binding energy---the property that the nuclear binding energy per nucleon
approaches a constant for a large number of nucleons.

\subsection{Numerical estimates}
\label{sec5}

Let us numerically estimate the nuclear binding energy. 
We fix the parameters of the model $\lambda$ and $M$ by using 
the mass of a nucleon and that of $\Delta$ as inputs. 

Together with the zero point fluctuation, 
the energy of 
the state with the isospin $I$ and the baryon number $A=1$ (a single baryon) in the nuclear matrix model \cite{Hashimoto:2019wmg} is 
given by 
\begin{equation}
 E_w = \left(N_c + 2 N_f\right) M + \frac{4\lambda}{M^2} I (I+1)
\end{equation}
where the first term is the zero point energy coming
from all $N_f$ flavors, 2 spins and both $w$ and $\bar w$. 
In order to distinguish the mass of nucleons and $\Delta$, 
we took first order perturbative correction which is the last term of \eqref{H}. 
By using $N_c=3$ and $N_f=2$, the nucleon mass $M_N$ and the $\Delta$ mass $M_\Delta$ are expressed as  
\begin{align}
 M_N &= M_0 + 7 M + \frac{3\lambda}{M^2} \ , 
 \label{MN}
 \\
 M_\Delta &= M_0 + 7 M + \frac{15\lambda}{M^2} \ , 
\end{align}
where $M_0$ is the bare tension of the baryon vertex, 
\begin{equation}
 M_0 = \frac{\lambda_\text{QCD} N_c M_{KK}}{3^3\pi} = \frac{\lambda N_c^2}{3 M^2} \ . 
 \label{M0}
\end{equation}
By using the experimental data, $M_N = 939\ \mathrm{[MeV]}$ and $M_\Delta = 1232\ \mathrm{[MeV]}$, 
the parameters of the nuclear matrix model are fixed as%
\footnote{
By using these inputs, we obtain $\lambda_\text{QCD} = 7.47$ and $M_{KK} = 277\ \mathrm{[MeV]}$. 
They are slightly different from those in other works but still of the same order. 
} 
\begin{align}
 \lambda &= 3.13 \times 10^5\ \mathrm{[MeV^3]} \ , 
 &
 M &= 113\ \mathrm{[MeV]} \ . 
 \label{fitlm}
\end{align}
Our holographic nuclear binding energy per nucleon \eqref{binres} is evaluated as\footnote{
As described in \cite{Hashimoto:2019wmg}, the numerical coefficient in \eqref{m^3} can vary in the range from $8 \lambda_r A$ to $48 \lambda_r A$, depending on the harmonic approximation methods of the commutator square potential in the nuclear matrix model (and in generic matrix models). Using the range, our result \eqref{resb} ranges from $B/A\simeq 9.66$ [MeV] to $B/A\simeq 17.5$ [MeV].
In addition, note that in the nuclear matrix model the electromagnetic force is not taken into account. 
} 
\begin{equation}
 \frac{B}{A} \simeq 9.66 \ \mathrm{[MeV]} \ . 
 \label{resb}
\end{equation}

In experimental data, 
the naive average of the nuclear binding energy per nucleon measured is known to be roughly $B/A \simeq 8$ [MeV].
In another
 fitting of the experimental binding energy data by the empirical Bethe-Weizs\"acker mass formula,
the coefficient of the volume term (the large-$A$ leading term) shows $B/A \simeq 16$ [MeV].
Our numerical estimate \eqref{resb} is close to 
the values of the experiments, which is a sufficiently good agreement 
as a nuclear model of holographic QCD.

%

\vspace{5mm}
\acknowledgments
We would like to thank Takeshi Morita for his collaboration 
at the early stage of this work. 
We also like to thank Hiroshi Suzuki and 
Hiroshi Toki for their suggestions on the methods.
This work was supported 
in part by MEXT/JSPS KAKENHI Grant No.~JP17H06462 and No.~JP20K03930.


\begin{appendix}

\section{Mass rescaling}

In this appendix, we study how the mass coefficient in the kinetic term of a generic 
non-relativistic 
action is scaled by quantum corrections to the total energy. This results in the scaling of
the coupling \eqref{lambdar}.

A non-relativistic action of a free particle is obtained by the non-relativistic limit of 
a relativistic worldline action as 
\begin{align}
 S 
 &= - m \int dt\,\sqrt{- \left(\dot x^\mu\right)^2}
 \notag\\
 &\simeq \int dt\, \left[- m + \frac{1}{2}m\left(\dot x^I\right)^2 
 + \cdots \right] \ ,
 \label{mnon}
\end{align}
where $\mu$ is the spacetime index.
To see the effect of the quantum corrections, 
it is convenient to introduce an einbein $e$ on the worldline, 
\begin{align}
 S 
 &= \int dt\, \left[ \frac{m}{2e} \left(\dot x^{\mu}\right)^2 - \frac{1}{2} m e \right] 
 \notag\\
 &= \int dt\, \left[ \frac{m}{2e} \left(\dot x^{I}\right)^2 
 - \frac{1}{2} m \left(e + e^{-1}\right)\right]  \ . 
\end{align}
Suppose we have a correction to the total energy from some other sector 
(such as $w$-sector in the nuclear matrix model).
It would effectively introduce an additional cosmological constant term $\delta m$
on the worldline as
\begin{align}
 S 
 &= \int dt\, \left[ \frac{m}{2e} \left(\dot x^{I}\right)^2 
 - \frac{1}{2} m \left(e + e^{-1}\right) - e \,\delta m\right] \, .
 \label{S+dm}
\end{align}
With this correction, the equation of motion for the einbein $e$ is solved as 
\begin{equation}
 e^2 = \frac{1- (\dot x^I)^2}{1 + (2 \delta m)/m} \ , 
 \label{e}
\end{equation}
in which $(\dot x^I)^2$ is negligible in the non-relativistic limit. 
Substituting this solution for the einbein, the corrected action
\eqref{S+dm} again takes the same form as \eqref{mnon}, 
but now the mass $m$ is replaced by 
 $M = m \sqrt{1 + (2 \delta m)/m}$. 
Thus, the correction to the total energy rescales the mass coefficient of the kinetic term 
as well as the cosmological constant on the worldline. 
This effect can also be interpreted as the rescaling of the einbein $e \simeq m/M$.

\begin{figure}[t]
\begin{center}
\includegraphics[width=7.5cm]{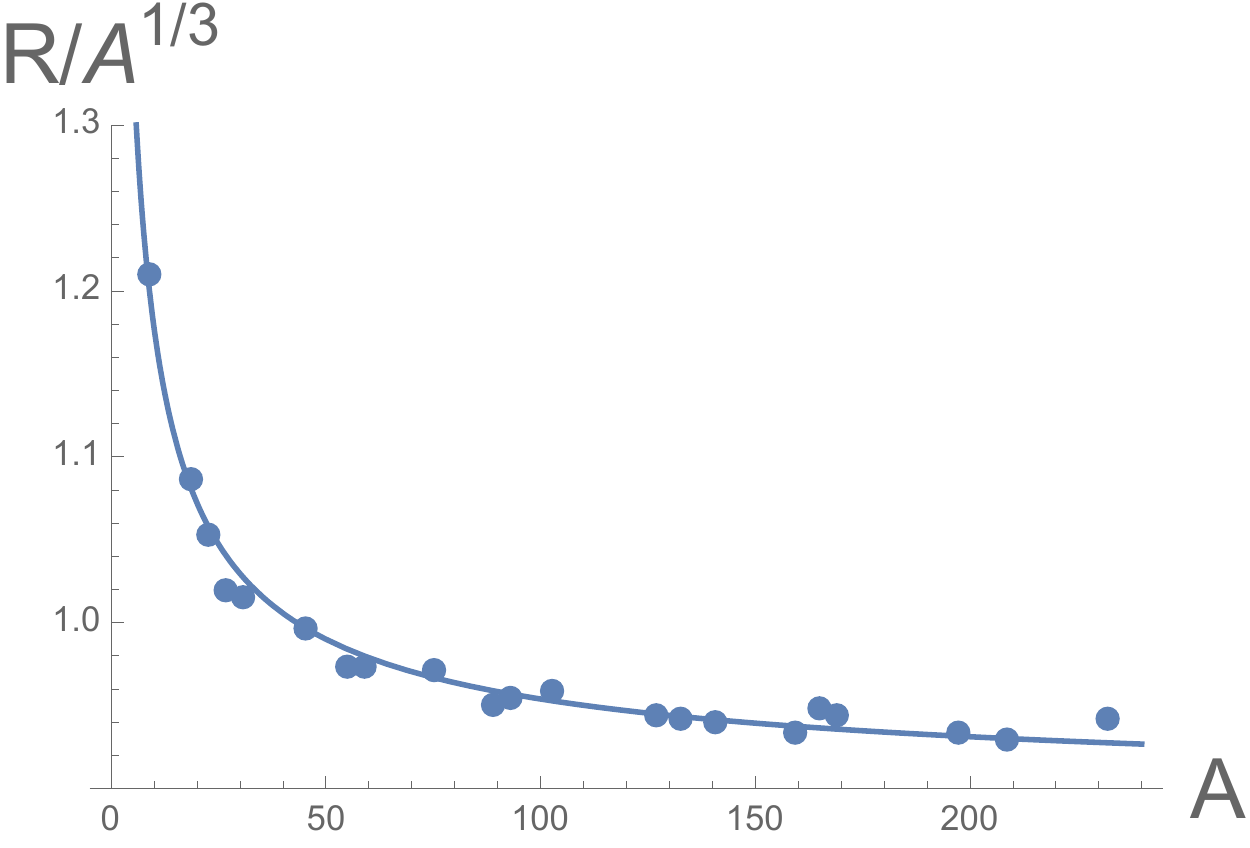}
\caption{
Experimental data \cite{table} of the charge radius $R$ of all known mononuclidic elements.
The solid line is the fitting curve \eqref{2/3} with the power $A^{-2/3}$ predicted by the holographic QCD. We find that the behavior is 
consistent.
\label{fig:A}}
\end{center}
\end{figure}

With this rescaling of the mass in mind, 
we consider the correction to the mass in the $X$-sector in the nuclear matrix model. 
Introducing the einbein, the action for $X^I$ is naturally written as
\begin{align}
 S 
 &= \int dt\, \tr\left[\frac{1}{2e} \left(D_t X^I\right)^2 
 + 2 e\,\lambda \left[X^I,X^J\right]^2 + \cdots \right] \ . 
 \label{A6}
\end{align}
The natural kinetic term can be given by a rescaling of $X^I$ to 
that in the physical length scale $X^I_\text{phys}$ as
\begin{align}
 S 
 &= \int dt\, \tr\left[\frac{M_0}{2e} \left(D_t X^I_\text{phys}\right)^2 \right.
\notag
\\
& \left. \quad 
 + 2 e\,M_0^2\lambda \left[X^I_\text{phys},X^J_\text{phys}\right]^2 + \cdots \right] \ , 
\end{align}
where the tension of the baryon vertex $M_0$ is given by \eqref{M0}. 
In this frame, we understand that the energy of $w$ gives the correction to the tension as \eqref{MN},
and modifies the einbein such that 
\begin{equation}
 e = \frac{M_0}{M_N} \ .
\end{equation}
Using this corrected $e$ and redefining the coordinate as 
\begin{equation}
 X^I= M_N^{1/2} X^I_\text{phys} \ ,  
\label{Xphys}
\end{equation}
we find that the action takes the same form as \eqref{A6}, 
\begin{align}
 S 
 &= \int dt\, \tr\left[\frac{1}{2} \left(D_t X^I\right)^2 
 + 2 \lambda_r \left[X^I,X^J\right]^2 + \cdots \right] \ ,
 \label{S(Xr)}
\end{align}
but with the rescaled coupling $\lambda_r\equiv e^3 \lambda$, which is 
our \eqref{lambdar}.

\section{$A$-dependence of nuclear radius}

With $m$ determined in \eqref{m^3}, our evaluation \eqref{<X^2>} shows 
that for the nuclear radii growing as $A^{1/3}$ there exists a sub-leading correction
in the large $A$ expansion. Expanding \eqref{<X^2>} for large $A$, we find that
the sub-leading correction starts at $A^{-2/3}$. This peculiar power is the result of
holographic QCD.

It is surprising that the nuclear experimental data indeed follow this $A^{-2/3}$ power law.
In Fig.~\ref{fig:A}, we plot the experimental data \cite{table} of the charge radius $R$ of all known mononuclidic elements.\footnote{The mononuclidic elements are 
21 chemical elements that are found naturally on Earth essentially as a single nuclide.
We use them because most of them are stable and were very-well measured with a high accuracy.} 
We fit the data with a function
\begin{align}
\frac{R}{A^{1/3}} = c_1 \left(1 + c_2 A^{-2/3}\right) \, 
\label{2/3}
\end{align}
and find a consistent fit with $c_1 = 0.89$ [fm] and $c_2 = 1.5$ (the solid line in Fig.~\ref{fig:A}). As this fitting curve reproduces the data very well, we conclude that
the sub-leading correction is given relatively by $A^{-2/3}$---the prediction of the
holographic QCD is confirmed in experiments.

We may even quantitatively compare the coefficients $c_1$ and $c_2$ with the nuclear matrix model.
Although in the model 
there could be some other sub-leading effect, here we simply assume the validity of
\eqref{<X^2>} and look at how the nuclear radius changes in $A$. 
Using \eqref{<X^2>} itself, the mean-square radius of the atomic nucleus in the nuclear matrix model is
\begin{align}
R_{\rm holo} & \equiv  
\sqrt{\frac{1}{A}\Big\langle\tr (X^I_{\rm phys})^2\Big\rangle} \ , 
\label{Rholo}
\end{align}
where 
$X^I_\text{phys}$ is related to $X^I$ in \eqref{<X^2>} by \eqref{Xphys}. 
Correcting \eqref{m^3} to the next-to-leading order, we have 
\begin{equation}
 m^3 = 8 \lambda_r A + 2^{5/3}3^{4/3} \lambda_r A^{1/3} \ .   
\end{equation}
Then, the coefficients $c_1$ and $c_2$ in the charge radius \eqref{Rholo} are calculated as 
\begin{align}
c_1 = \frac{3^{1/2}A^{1/3}}{2\lambda_r^{1/6} M_N^{1/2}} = \frac{M}{2\lambda^{2/3}} \, , \quad 
c_2 = \frac{3^{1/3}}{2^{4/3}} \, .
\end{align}
Using the numerical parameters \eqref{fitlm} of the nuclear matrix model, 
we find $c_1 \simeq 2.4$ [fm] and $c_2 \simeq 0.57$.
These values are of the same order as those of the experiments, which is 
quite a good agreement in view of the crude approximations in the holographic QCD.

\end{appendix}


\end{document}